\documentclass{ifacconf}
\usepackage{subfigure}
\usepackage{graphicx}

\usepackage{color}
\usepackage{times}
\usepackage{stfloats}    
\usepackage{natbib}        
\usepackage{amsmath}
\begin{document}
\begin{frontmatter}

\title{Perceived Time To Collision as Public Space Users' Discomfort Metric\thanksref{footnoteinfo}} 

\thanks[footnoteinfo]{This work was supported by the Ministry of Science and Technology
(MOST), Taiwan, under Grant MOST 110-2636-E-006-005 and MOST 111-
2636-E-006-004, and the Higher Education Sprout Project, Ministry of
Education to the Headquarters of University Advancement at National Cheng
Kung University (NCKU).}

\author[First]{Alireza Jafari} 
\author[First]{Yen-Chen Liu} 
\address[First]{Networked Robotics System Laboratory, Department of Mechanical Engineering, National Cheng Kung University, Tainan, Taiwan, (alirezajafari110@gmail.com and yliu@mail.ncku.edu.tw).}

\begin{abstract}                
Micro-mobility transport vehicles such as e-scooters are joining current sidewalk users and affect the safety and comfort of pedestrians as primary sidewalk users. 
The lack of agreed-upon metrics to quantify people's discomfort hinders shared public space safety research.
We introduce perceived Time To Collision (TTC) as a potential metric of user discomfort performing controlled experiments using an e-scooter and a pedestrian moving in a hallway. 
The results strongly correlate the participant's reported discomfort and the perceived TTC.
Therefore, TTC is a potential metric for public space users' discomfort. Since the metric only uses relative velocity and position information, it is a viable candidate for neighboring people's discomfort estimation in  advanced driver assistance systems for e-scooters and PMVs.
Our ongoing research extends the results to mobile robots. 
\end{abstract}


\end{frontmatter}
\section{Introduction}
Micro-mobility transport vehicles are getting more popular in recent years.
Examples are e-scooters, autonomous delivery robots, and even intelligent wheelchairs; see~\cite{T-ITS-I, E-BikeBehavior,RobotExample,WhCExmple}.
Feeling vulnerable, these new road users are not comfortable moving side by side with the cars in the streets.
Thus, they join sidewalks making pedestrians, the primary sidewalk users, uncomfortable. 
Therefore, studying the new users' integration and interactions with pedestrians helps improve everyone's safety and comfort.
\par
Despite the recent integration and its imposed challenges, research on the social acceptability of these modern sidewalk users is scarce.
\cite{Rob_Nav_2022} reviewed the social acceptability of mobile robots and pointed out the lack of agreed-upon benchmarks in evaluating social acceptability.
The lack of metrics in public spaces leads to arbitrary policy-making and infrastructure designs endangering public safety and comfort.
Therefore, the discomfort metrics of a human, either as a pedestrian or as a Personal Mobility Vehicle (PMV) user, are necessary for further integration of new users into public spaces.
\par
The autonomous mobile robots' presence in public spaces is a pressing challenge, too.
Adding mobile robots into our shared public spaces requires them not to raise undue discomfort. 
Yet, detecting these feelings is crucial for proactively caring for and addressing people's feelings.
The mobile robots' awareness of surrounding pedestrians helps them to make more socially acceptable decisions; see~\cite{Saf_Con, App_Ped}.
Namely, the people's discomfort integration into the robots' decision-making algorithms can potentially improve pedestrian comfort. 
The first step toward this goal is quantifying human discomfort. 
According to \cite{Saf_Con}, psychological discomfort arises from the agent's appearance and movement.
As the primary movement control states, this research uses relative position and velocity to estimate discomfort --- a prerequisite for socially aware mobile robot design.
\par
So far, due to the lack of measurements, social robotics research has focused on the open-loop improvement of human feelings. 
For example,~\cite{Shiomi2014} suggested that if the mobile robot's movement patterns are human-like, the neighboring pedestrians will perceive it as natural and, Thus, feel more comfortable.
Their mobile robot employed Social Force Model (SFM) to imitate human collision avoidance patterns in a shopping mall.
As a result, the human experience around the mobile robot improved without real-time knowledge of people's discomfort---hence open-loop.
The estimation of the discomfort using position and velocity states can fill the gap and close the feedback loop in the controller design. 
It enables the robot's control algorithm to understand neighboring pedestrians' discomfort and care for them.
\par
\cite{Hasegawa2018} experimentally studied pedestrians' danger perception toward PMVs using a pedestrian and a Segway-type PMV interacting in a controlled hallway.
They recorded the pedestrian's subjective danger during the trials using questionnaires and compared it with the estimated danger using an SFM-based model. 
They plotted the observed danger versus the estimated one.
Judging by the coefficient of determination $R^2\approx0.68$, SFM acted as an acceptable subjective danger estimator.  
Moreover,~\cite{T-ITS-I} used accumulated acceleration and social force as pedestrian safety and comfort metrics during Monte Carlo simulations.
Inspired by the potential of closing the feedback loop, this research experimentally studies the relation between the reported discomfort and the perceived Time To Collision (TTC) as a real-time estimator/indicator of neighboring pedestrians' feelings around e-scooters.
\par
Originally, TTC appeared in transportation safety research to determine near-miss car traffic events; see~\cite{FORMOSA}
\cite{Hayward1972} used trained but unaware human observers to watch video recordings of dangerous events and then collected their opinion on the danger levels.
The minimum TTC successfully predicted most of the perceived danger levels by the observers. 
They also reported that most car drivers maintain a constant TTC to the target car, which aligns with our findings of an e-scooter interaction with a pedestrian.
\par
Years later, traffic safety research still recognizes TTC as an important measure of an accident's imminence.
\cite{Archer2005} named the minimum TTC as a conflicts' primary time-based severity surrogate measure.
Later, case-specialized definitions of TTC emerged.
\cite{Zhang2012} addressed less studied pedestrian-involved collisions with cars focusing on a variant of TTC.
They defined the Time Difference To Collision (TDTC) and evaluated its performance in classifying dangerous situations using recorded video data of actual zebra crossing.
In addition, \cite{Zhang2022} developed an algorithm for collision risk assessment of a car facing multiple pedestrians. 
They calculated a potential collision area using linear and angular velocities of both vehicles and pedestrians.
The method selects the pedestrian with minimum TTC to the area as the most vulnerable target.
\par
Moreover, many researchers focused on the practicalities of applying TTC to actual vehicles.
For example, \cite{Jiang2015} used TTC to address the conflict moments' behavioral differences between driving cultures in Beijing and Munich.
These differences are essential in Advanced Driver Assistance System's (ADAS) adaptations to various cultures.
Another example is a Deceleration-based Surrogate Safety Measure (DSSM) that considers both mechanical braking capability and driver braking behavior; see~\cite{Tak2015}.
In addition, \cite{Kilicarslan2018} focused on extracting TTC as an indicator of dangerous events from a single camera installed on a vehicle as an alternative for the depth sensing required for calculating TTC.
\par
Recent advances in Vehicle-to-vehicle (V2v) and Vehicle-to-Infrastructure (V2I) communications open new fronts for TTC applications. 
\cite{Jang2012} proposed a cooperative collision warning system in unsignalized intersections when the approaching vehicles can not see each other. 
They used the information received from the other car to calculate TTC continuously and generate an alarm accordingly.
Moreover,~\cite{Jo2022} developed an in-vehicle forward collision warning system to predict crash risks in real time using vehicle interaction data obtained through V2V communications.
Instead of just using the sensors' trajectory data, their algorithm employs the collected information from other vehicles and infrastructure to calculate a real-time crash risk as a function of TTC and share it with other Connected Vehicles (CV).
\par
Robotics research also used TTC to indicate the possibility of a crash.
For instance, \cite{Masaki2020} formulated force feedback to an operator remotely controlling a mobile robot. 
The controller provided force assistance to the operator in a bilateral tele-operative manner as a function of TTC---a metric of a potential crash. 
Moreover, \cite{Shahriari2022} developed a navigation algorithm for heterogeneous mobile robots using TTC.
Since the methods relying on only kinematics may underestimate or overestimate TTC, they used dynamic models into the motion control algorithm.
In doing so, they incorporated TTC and its time derivative into their nonlinear controller and proved that the system is stable in the sense of Lyapunov.
\par
This study proposes TTC as a metric for public space discomfort.
Defining perceived TTC as a variant of TTC, we performed controlled experiments with an e-scooter and a pedestrian interacting with each other and continuously recorded TTC. 
Moreover, we recorded participants' reported discomfort when facing and passing each other.
The results show a strong correlation between the reported discomfort and the minimum TTC during the trials when a participant, either the e-scooter rider or the pedestrian, sees the other.
Thus, we believe TTC can estimate people's discomfort in shared public spaces.
\par
\section{Perceived Time to collision}
\begin{figure}
\begin{center}
\includegraphics[width=0.48\textwidth]{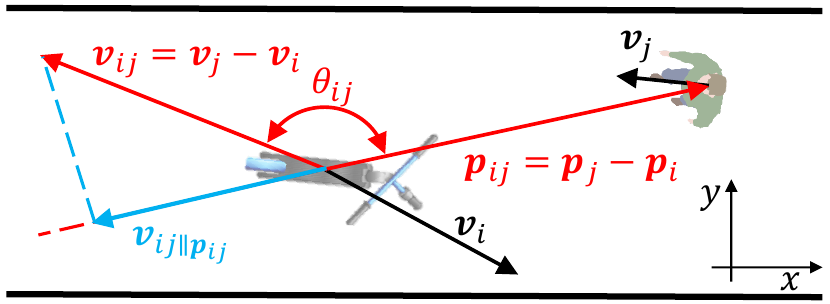}   
\caption{The perceived approach rate $\vec{v}_{ij \parallel \vec{p}_{ij}}$ definition for agents modeled as centered masses.} 
\label{fig:model}
\end{center}
\end{figure}
This section introduces the perceived time to collision by extending TTC to 2-dimensional space. 
By definition, time to collision is the remaining time to an impact between two approaching objects on a straight line.
Assuming constant velocities, centered masses, and moving in 1-dimensional space, the simplest formulation of the time to collision is
\begin{equation} \label{TTCBasic}
TTC=\frac{\Delta X}{\Delta V},
\end{equation}
where $\Delta X$ and $\Delta V$ are the relative distance and velocity between the two objects/vehicles.
\par
When two agents, such as vehicles, pedestrians, or e-scooters, interact, their behavior depends on how they perceive each other rather than how they actually move.
For example, in a 2-dimensional space, the two approaching agents may never theoretically collide because their trajectories are skew lines in $x$-$y$-$t$ space.
However, their behavior is affected by the rate they seemingly approach each other and perceive potential collisions.
\par
Fig.~\ref{fig:model} is a shared space with two agents modeled as centered masses, e.g., an e-scooter and a pedestrian.
$\vec{p}_{i}$, $\vec{p}_j$, $\vec{v}_i$, and $\vec{v}_j$ are their positions and velocity vectors in the local coordinate frame~$x$-$y$, respectively; subscripts $i$ and $j$ denote e-scooter and pedestrian.
The relative position and velocity of agent $j$ with respect to agent $i$ are $\vec{p_{ij}}$ and $\vec{v_{ij}}$,
\begin{equation} \label{RelativePosition}
\vec{p_{ij}}=\vec{p}_j-\vec{p}_i
\end{equation}
and
\begin{equation} \label{RelativeVelocity}
\vec{v_{ij}}=\vec{v}_j-\vec{v}_i.
\end{equation}
The agents' perception of time to collision depends on the rate they get closer to each other in the line of sight, $\frac{d}{dt}\|\vec{p}_{ij}\|$.
Therefore, the obtained vector $\vec{v}_{ij \parallel \vec{p}_{ij}}$ from projecting the relative velocity vector, \eqref{RelativeVelocity}, onto the relative position vector, \eqref{RelativePosition}, is the perceived approach rate,
\begin{equation} \label{ApproachRate}
\frac{d}{dt}\|\vec{p}_{ij}\|=\|\vec{v}_{ij \parallel \vec{p}_{ij}}\|=\frac{{\vec{p}_{ij}}.{\vec{v}_{ij}}}{\|{\vec{p}_{ij}}\|}
\end{equation}
where "$.$" is the vector's inner product.
Following \eqref{TTCBasic}, the perceived TTC, $T_p$, in a 2-dimensional space is
\begin{equation} \label{PTTC}
T_p=\frac{{\|{\vec{p}_{ij}}\|}^2}{{\vec{p}_{ij}}.{\vec{v}_{ij}}}.
\end{equation}
In addition, since 
\begin{equation}
{\vec{p}_{ij}}.{\vec{v}_{ij}}={\|{\vec{p}_{ij}}\|}{\|{\vec{v}_{ij}}\|}cos\theta_{ij},
\end{equation}
an alternative formulation of $T_p$ is
\begin{equation} \label{PTTCSimple}
T_p=\frac{{\|{\vec{p}_{ij}}\|}}{{\|{\vec{v}_{ij}}\|}cos\theta_{ij}}
\end{equation}
where $\theta_{ij}$ is the angle between the relative position and velocity. 
Note that $T_p$ is bilateral and has the same value for both interacting agents.\par
\section{Design of Experiments}
\begin{figure}
\centering
$
\begin{array}{ccc}
\subfigure[] \centering {\includegraphics[width=0.155\textwidth]{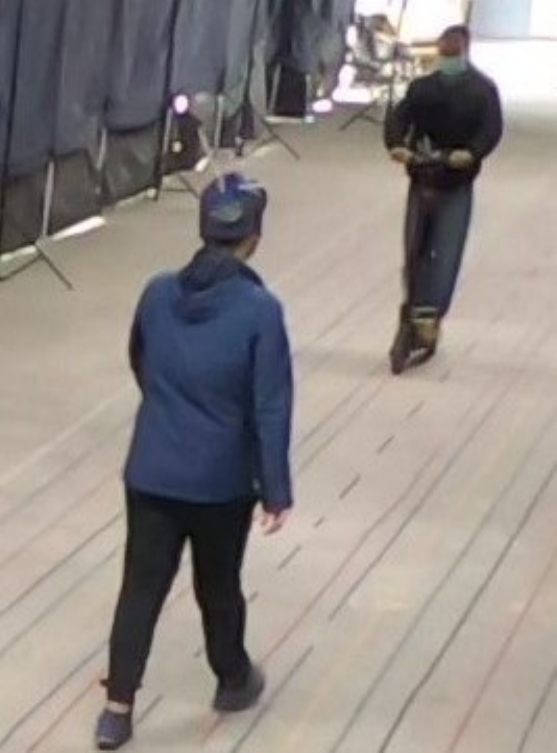}}  & \hfill
\subfigure[] \centering {\includegraphics[width=0.16\textwidth]{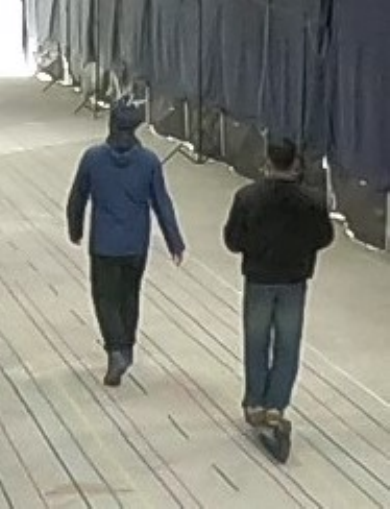}}  & \hfill
\subfigure[] \centering {\includegraphics[width=0.13\textwidth]{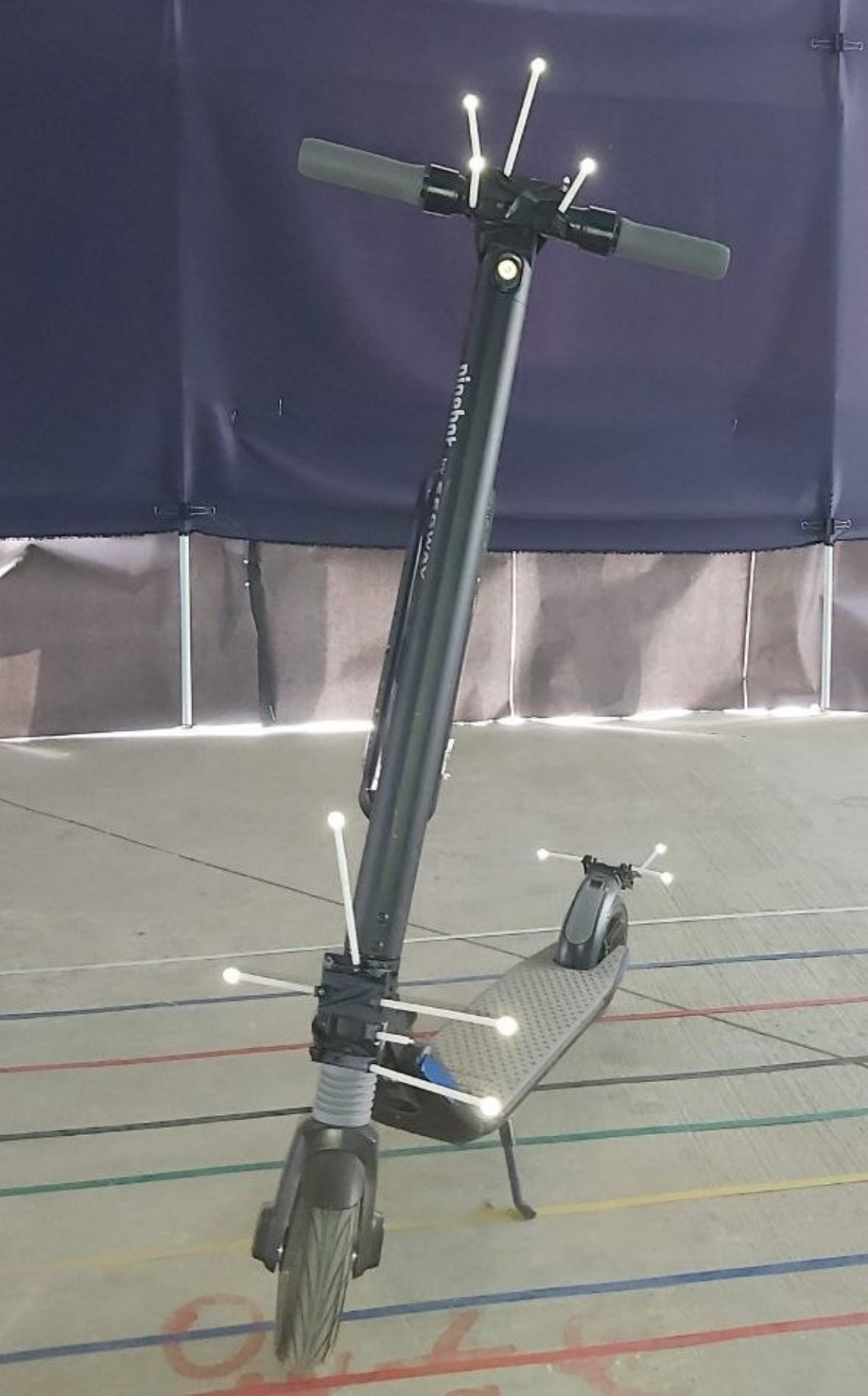}} 
\end{array}$
  \caption{
(a) An e-scooter is facing a pedestrian in a controlled sidewalk;
(b) An e-scooter is passing a pedestrian in a controlled sidewalk;
(c) The retro-reflective markers installed on the e-scooter.
}\label{fig:exp}
\end{figure}
\begin{figure}
\begin{center}
\includegraphics[width=0.48\textwidth]{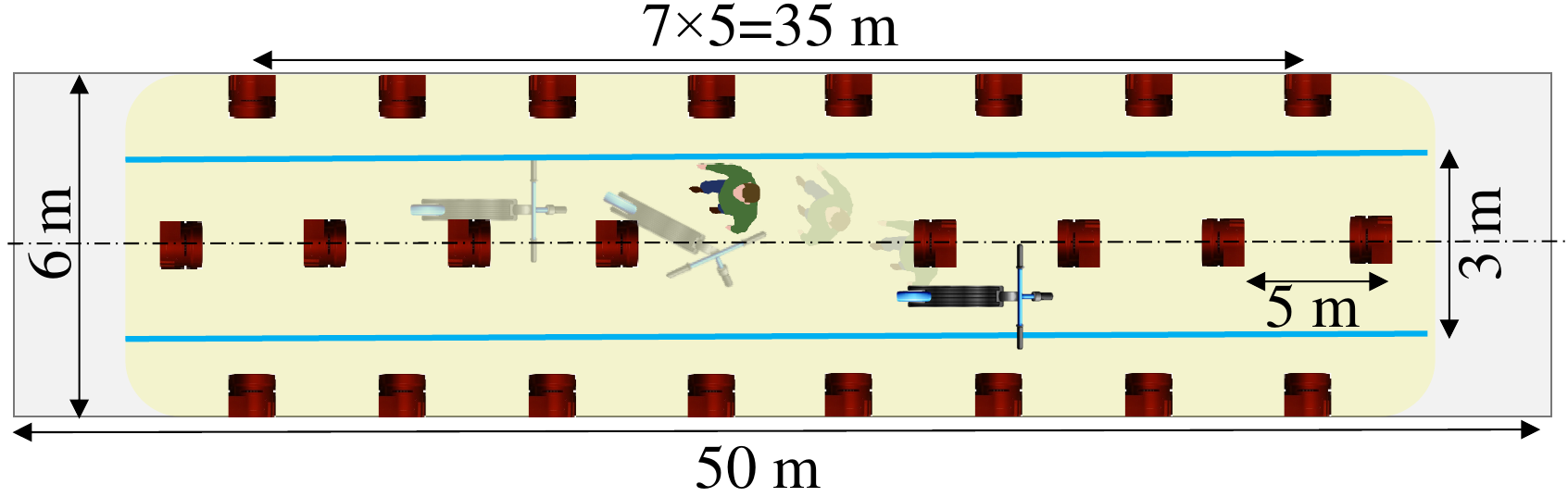}   
\caption{Experiments' hallway and cameras arrangement.} 
\label{fig:exparea}
\end{center}
\end{figure}
The experiments' trials cover two main situations: facing trials and passing trials. 
The facing trials are the trials with an e-scooter rider moving toward a pedestrian inside a bounded sidewalk and passing him/her, as shown in Fig.~\ref{fig:exp}(a).
During the passing trials, the e-scooter rider approaches the pedestrian from behind on the same sidewalk and overtakes him/her, as shown in Fig.~\ref{fig:exp}(b).
We repeated each of the passing and facing trials for ten different participants' sets (a set is a pedestrian and an e-scooter rider) and ten times per set.
After each trial, each participant filled out a questionnaire reporting his/her maximum discomfort level.
In total, we collected 400 data points from 2$\times$100 trials, facing and passing, and two participants per trial.
The subjects were instructed to report their discomfort using seven choices: 0-Comfortable, 1-Slightly uncomfortable, 2-Uncomfortable, 3-Annoying, 4-Dangerous, 5-Very dangerous or near crash, and 6-Collision.
The description for each number helps reduce the personal interpretation of the safety levels.
\par
All participants were 20 to 35 years old and had less than six months of e-scooter riding experience. 
Before the trials, they had the chance to ride around the experiment area and get familiar with the environment.
They knew that the research goal was to study e-scooter riders' behavior in general. However, they were unaware of specific research goals.
\par
The e-scooter used in the experiments is an ES4 Ninebot Kick scooter from Segway.
The model's maximum velocity is 8.3 m/s, yet, out of concern for participants, a top speed of 5 m/s was forced in the e-scooter's internal setting. 
In addition, we instructed the rider to move at a comfortable speed.
\par
The hallway dimensions are 6 m $\times$ 50 m, but we asked the participants to stay inside a three meters wide area marked on the ground as virtual sidewalk curbs.
Moreover, the cameras targeted the central 35 meters for data collection. The rest of the hallway was for acceleration and deceleration.
We selected the width because it is not too wide to let the participants pass each other without any significant interaction, and it is not too narrow to affect the participants' feelings and obscure their judgment of discomfort caused by the other participant.
\par
We installed multiple groups of retro-reflective markers on the pedestrians' helmets, Fig.~\ref{fig:exp}(a) and Fig.~\ref{fig:exp}(b), and on the scooter, Fig.~\ref{fig:exp}(c).
As arranged in Fig.~\ref{fig:exparea}, a motion capture system with 24 infra-red cameras (Optitrack Flex 13) sees the markers' reflection and locates the participants. 
Sixteen cameras installed on the hallway's walls cover the central area. 
The eight cameras on the ceiling provide redundancy in marker detection to compensate for blind spots and partial blockage of the cameras' line of sight by participants' bodies.
\par
The data-processing software (Motive) records the timed trajectories of the participants at 120 Hz.
The recorded timed trajectories are used to calculate relative positions and velocities. 
We calculated $T_p$ using these relative states at each moment. 
Note that during both types of trials, before the passing point and when the two agents are approaching each other, $T_p$ is positive.
Thus, we selected the minimum value during each trial before the passing moment as the perceived TTC. 
This paper studies the relation between this minimum perceived time to collision and shared space users' discomfort.
Therefore, from now on, the term "perceived time to collision" refers to the minimum $T_p$ during each trial.
\par
\section{results and discussion}
This section presents and discusses the results collected from the experiments. First, we focus on the recorded $T_p$ and then discuss the reported discomfort. 
Finally, we present the relationship between the two.
\subsection{Percieved TTC for Passing/Facing Cases}
\begin{figure}[t!]
\begin{center}
\includegraphics[width=0.48\textwidth]{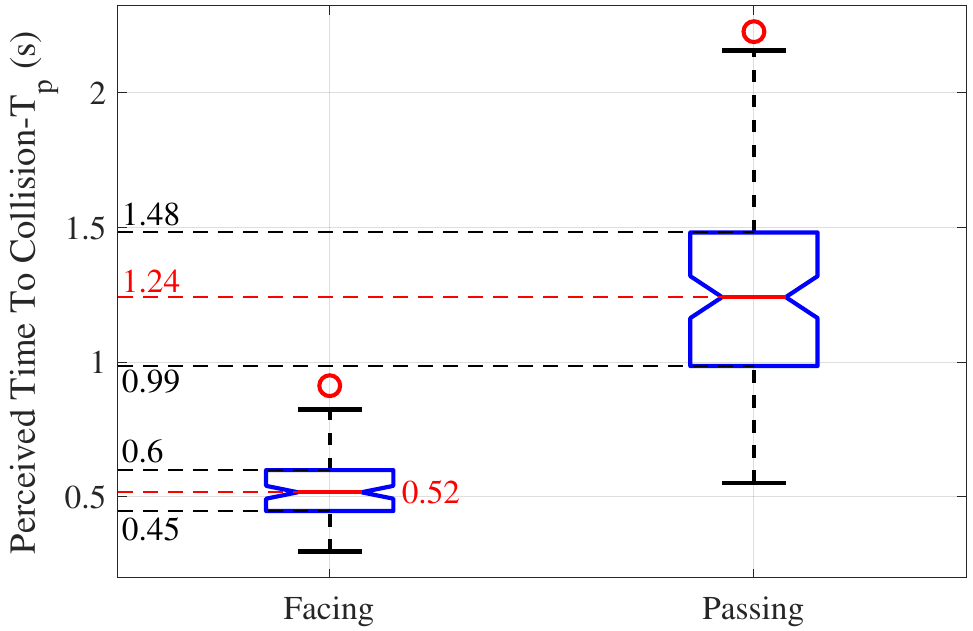}   
\caption{Perceived TTC for passing and facing trials. The median and upper/lower quartiles' edges are marked by red and black dashed lines.} 
\label{fig:TTCBarChart}
\end{center}
\end{figure}

Fig.~\ref{fig:TTCBarChart} shows box plots of the perceived time to collisions for facing and passing trials. 
The boxes contain the central quartiles, and the whiskers represent the upper and lower quartiles.
The values considerably far from the rest of the group, or outliers, are marked by “o”. 
The narrowing around the median, or the notch, is an indicator for statistical comparison. Usually, not overlapping notches are evidence of a significant difference between the two data groups. 
For more detail see \cite{BoxChart}.
\par
Fig.~\ref{fig:TTCBarChart} shows a significant behavior difference between the facing and passing trials. 
The perceived time to collision for the facing case for all participants is 0.52 s, while for the passing case, it is much higher, 1.24 s.
An explanation is that when the e-scooter rider approaches the pedestrian from behind, the rider maintains a larger response time to cover possible sudden movements of the pedestrian. 
However, they don't expect sudden dangerous movements when they face and see each other.
Therefore, they move faster and maintain a shorter time to collision.
Note that for the facing case, the recorded time to collision is the result of cooperation between the e-scooter rider and the pedestrian, whereas in the passing case, because the pedestrian can not see the rider, it only shows the behavior of the rider.
\par
\begin{figure}
\begin{center}
\includegraphics[width=0.48\textwidth]{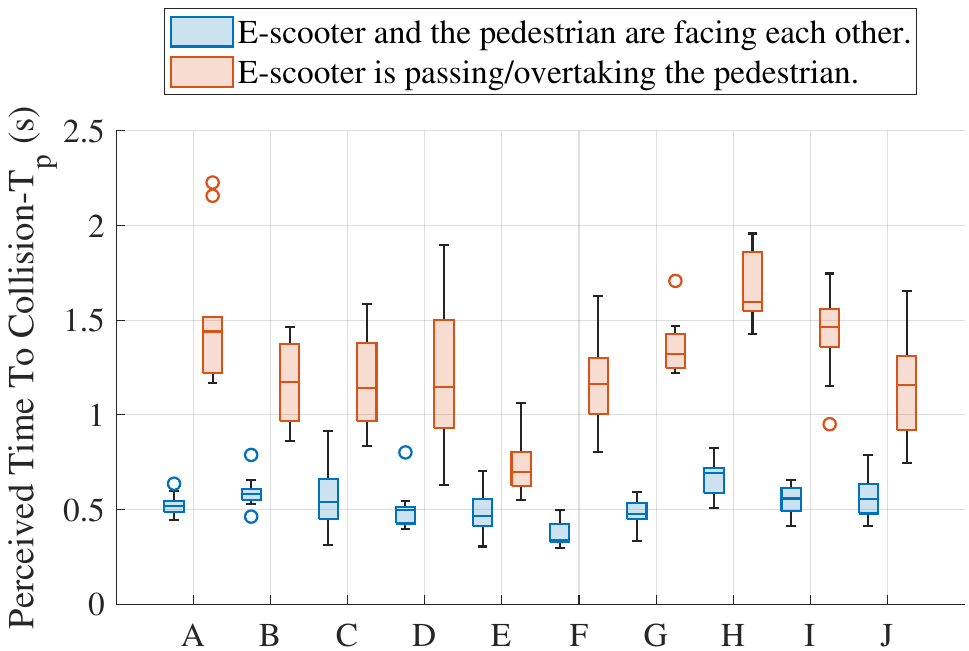}    
\caption{Perceived TTC by individuals during passing and facing trials. The circles and lines are the outliers and the medians, respectively. } 
\label{fig:TTCPeopleBarChart}
\end{center}
\end{figure}

The box plots in Fig.~\ref{fig:TTCPeopleBarChart} are TTCs of the riders marked by capital letters A to J. 
The figure presents the facing cases' perceived TTCs (blue) and the passing cases' perceived TTCs (red).
Fig.~\ref{fig:TTCPeopleBarChart} shows that the mentioned observation applies to all participants at an individual level, too. 
Thus, when the e-scooter rider approaches the pedestrian from behind, the perceived TTC is much higher independent of the rider's behavioral differences.
Another observation is that most e-scooter drivers maintain an almost constant TTC when interacting with a pedestrian.
\cite{Hayward1972} also reported this behavior for car drivers in near-miss traffic events. 
Additionally, during the facing trial, when the participants see each other and the actions are cooperative, the range of the recorded minimum perceived time to collisions is much narrower than the passing case when the action is just a result of the rider's behaviors. It suggests that the rider limits its response range when cooperating with other agents, perhaps due to a repetitively learned social pattern.
Further research is required to study the correlation.
\par
\subsection{Observed Discomfort for Passing/Facing Cases}
\begin{figure}[t!]
\begin{center}
\includegraphics[width=0.48\textwidth]{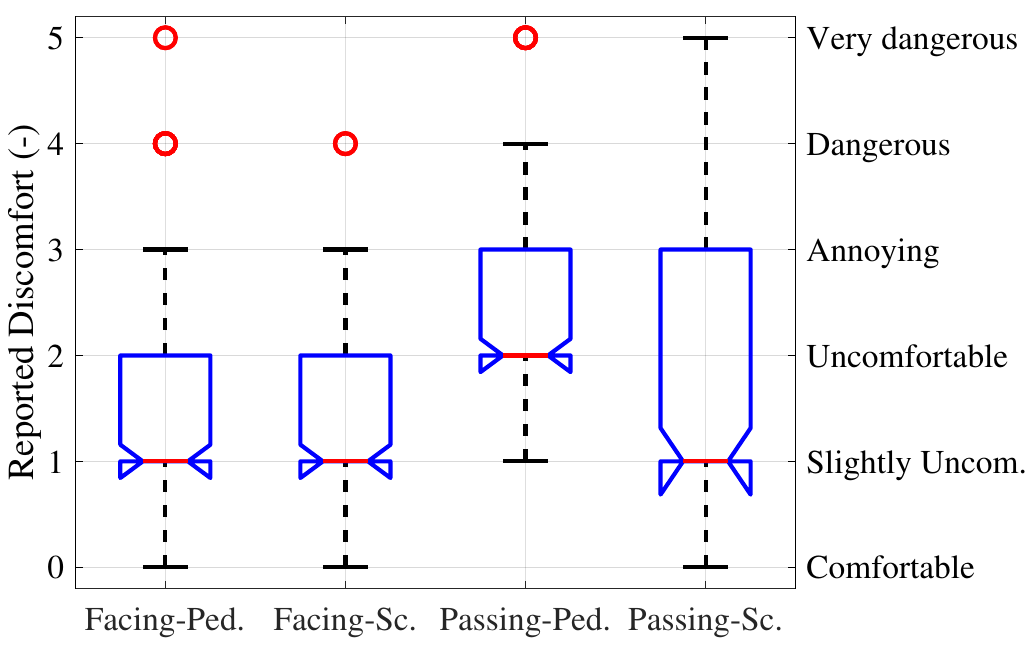}   
\caption{Reported discomfort for the pedestrian and the rider during facing and passing trials. Red circles and red line are the outliers and the medians, respectively.} 
\label{fig:RDBarChart}
\end{center}
\end{figure}
This section focuses on the reported discomfort level during the facing and passing trials for the pedestrian and the e-scooter rider. 
Fig.~\ref{fig:RDBarChart} shows the box plots of the reported discomfort of both subjects during the facing and passing trials.
For the e-scooter rider, the reported discomfort levels are not different between facing and passing trials despite the significant difference in the perceived TTC, i.e., Fig.~\ref{fig:TTCBarChart}. 
The reported discomfort levels are similar since the median values are the same and the notches overlap, ranging from slightly uncomfortable to uncomfortable. 
The similarity in reported discomfort and the significant difference in perceived TTC suggest that the riders adjust $T_p$ to guarantee a maximum bound on the discomfort. 
In other words, when the e-scooter rider feels slightly uncomfortable to uncomfortable, he takes actions, either steering away or braking, to increase the perceived time to collision.
The e-scooter rider takes actions to increase the perceived time to collision when it reaches approximately 0.5 s and 1.2 s for the facing and passing cases, respectively.
\par
On the other hand, for pedestrians, the reported discomfort significantly differs between facing and passing trials. 
The passing trials' median and the notch are higher than the facing cases showing higher reported discomfort ranging from uncomfortable to annoying compared to the facing case. 
During the facing trials, the pedestrian and the rider see each other and cooperatively take action to avoid violating their discomfort upper bound. 
However, since only the scooter rider sees the pedestrian during the passing trials, while the rider takes action by increasing the perceived time to collision as discussed, the pedestrian has no understanding of TTC.
The lack of knowledge prevents cooperative maintenance of the discomfort level. 
Thus, only the e-scooter rider's effort contributes to the pedestrian's comfort; therefore, the pedestrian's reported discomfort is higher during the passing trials than the facing trials.
\par
Next, we study the relationship between the reported discomfort level and the perceived time to collision and show that when the participants, either the pedestrian or the e-scooter rider, see the other one, there is a strong correlation. However, when they don't see each other, there is no relation, as expected.
\subsection{Observed Discomfort and Perceived TTC}
This section studies the relationship between the reported discomfort and perceived time to collision. 
We fit curves to the collected data and compare their $R^2$ as a goodness of fit measure.
\par
Briefly, $R^2$ or the coefficient of determination evaluates how well a curve fits a set of actual data points; if the data mean value is used as the predictor, $R^2=0$ and for a perfect predictor, $R^2=1$.
Thus, $R^2$ shows how much a fit is better than simple averaging.
For the definition of $R^2$, see ~\cite{RSquared}.
\par
\begin{figure*}[ht]
\centering
$
\begin{array}{cc}
\subfigure[] {\includegraphics[width=0.48\textwidth]{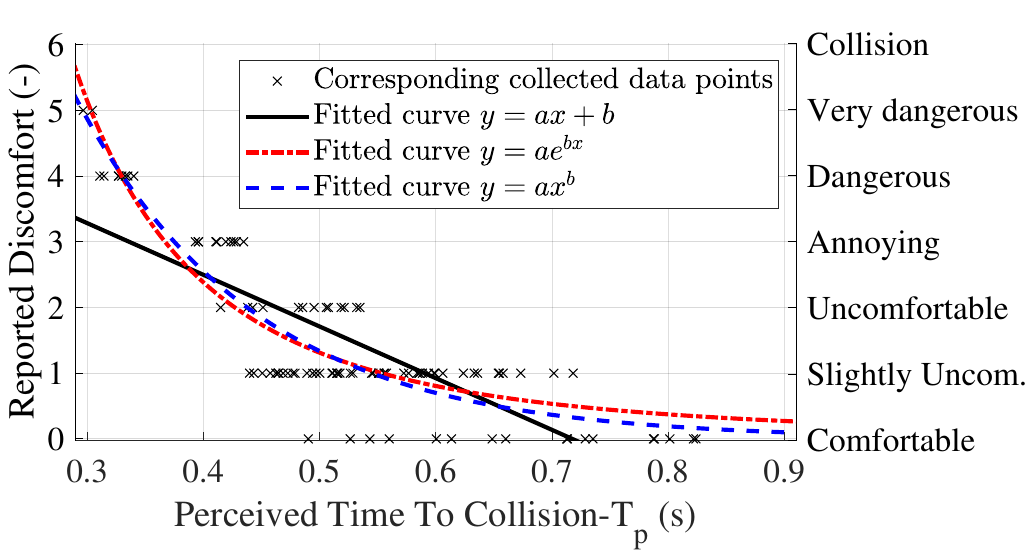}} &
\subfigure[]{\includegraphics[width=0.48\textwidth]{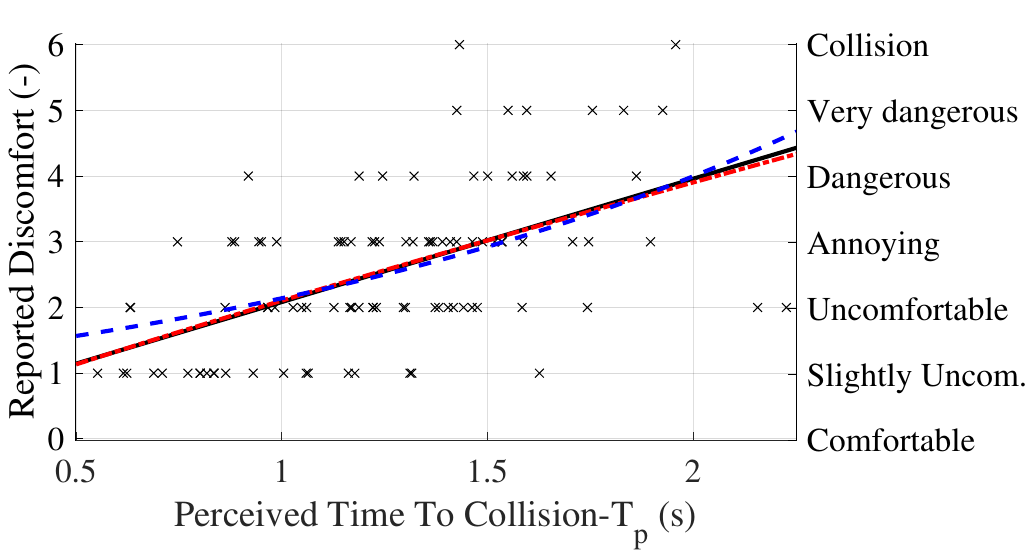}}  \\
\subfigure[] {\includegraphics[width=0.48\textwidth]{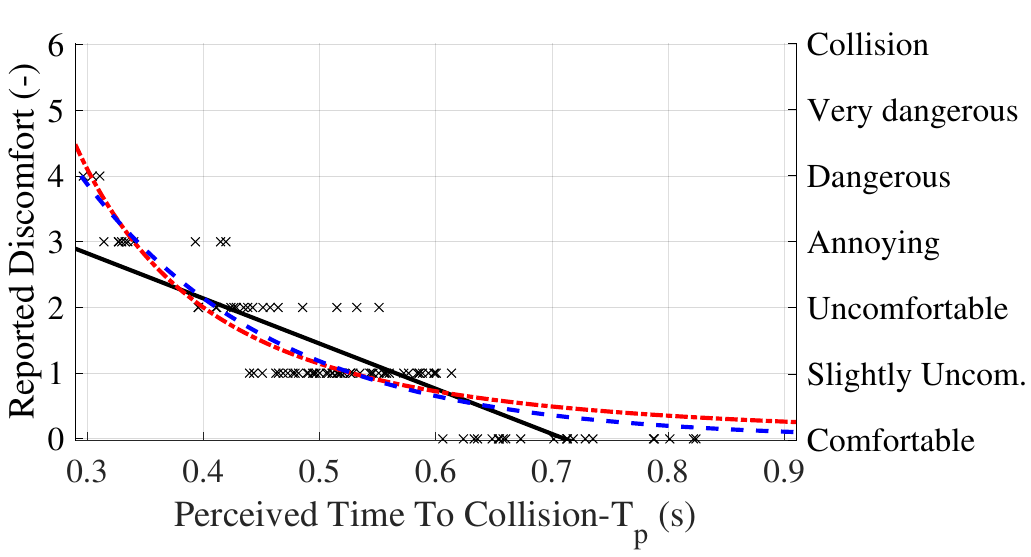}}  &
\subfigure[]{\includegraphics[width=0.48\textwidth]{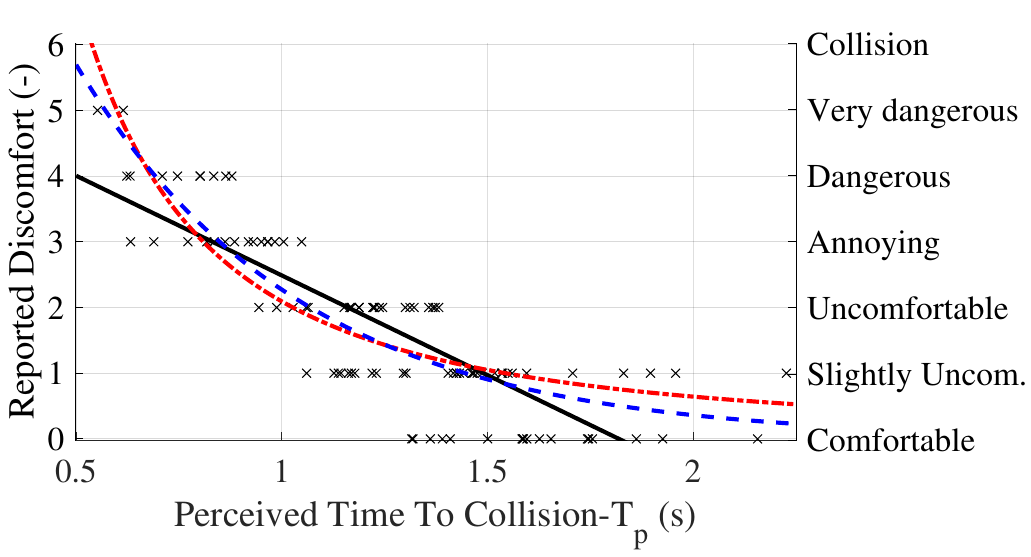}}  
\end{array}$
  \caption{Fitted functions to estimate the reported discomfort using perceived TTC for
(a) a pedestrian facing an e-scooter;
(b) a pedestrian when an e-scooter is approaching from behind;
(c) an e-scooter rider facing a pedestrian;
(d) an e-scooter rider approaching a pedestrian from behind.
Table~\ref{Table:Goodness of fit} presents the coefficient of determination and the constants. The legend applies to all.
}\label{fig:RDVsTTC}
\end{figure*}

Fig.~\ref{fig:RDVsTTC} presents the collected data points. 
Fig.~\ref{fig:RDVsTTC}(a) and Fig.~\ref{fig:RDVsTTC}(b) show the data points collected from the pedestrians during the facing and passing trials, respectively. 
Similarly, Fig.~\ref{fig:RDVsTTC}(c) and Fig.~\ref{fig:RDVsTTC}(d) contain the rider's data points during the facing and passing trials.
\par
For each case, we fitted two types of functions with two constants per function for a fair evaluation.
The candidates must satisfy two observed human behaviors. 
The first one is that we don't feel any discomfort when the time to collision is big enough. 
On the other hand, when the time to collision is short enough, we feel extreme discomfort.
Therefore, the functions should converge to zero when the perceived TTC is going to infinity, and they should have significantly large values when the perceived TTC approaches zero.
We selected exponential and power functions in the forms presented in Table~\ref{Table:Goodness of fit}. Moreover, we fitted lines as a base for comparison, too.
Table~\ref{Table:Goodness of fit} presents the functions, their coefficient of determination or $R^2$, and the obtained constants for each case.
Note that the parameters may vary with external conditions such as demographics, cultural differences, and user experience.
\par
Comparing the $R^2$ values, for the cases that the participant can see the other one, i.e., Fig.~\ref{fig:RDVsTTC}(a), Fig.~\ref{fig:RDVsTTC}(c) and Fig.\ref{fig:RDVsTTC}(d), a statistically significant relationship between the reported discomfort and perceived TTC exists.
Considering the existing randomness in human behavior, the goodness of fit for both estimators is significant.
Moreover, for the same reason, the difference between the fitted curves is not meaningful and possibly accidental.
Since the coefficient of determination difference between the candidates is not meaningful, both are suitable for discomfort estimations.
\par
However, in Fig.~\ref{fig:RDVsTTC}(b) where the pedestrian can not see the e-scooter approaching from behind, the relation is not significant; $R^2\le0.3$ generally indicates none or weak correlation.
We expected this because the pedestrian can not see the e-scooter and doesn't understand the TTC and, therefore, can not have reported discomforts related to the time to collision.
\par
In summary, we can estimate pedestrian and PMV rider discomfort using the fitted functions of perceived time to collision in real-time.
\begin{table}
\begin{center}
\caption{$R^2$ as the goodness of fit's measure and obtained constants for the fitted curves in Fig.~\ref{fig:RDVsTTC}}\label{Table:Goodness of fit}
\begin{tabular}{ccccccc}
Figure                & Fit type   & Equation    & $R^2$   & a      & b      \\\hline
\ref{fig:RDVsTTC}(a)  & Line       & $y=ax+b$    &  0.65   & -7.9   & 5.6     \\
\ref{fig:RDVsTTC}(a)  & Exponential& $y=ae^{bx}$ &  0.82   & 33.9   & -6.5    \\
\ref{fig:RDVsTTC}(a)  & Power      & $y=ax^{b}$  &  0.81   & 0.21   & -2.7    \\
\ref{fig:RDVsTTC}(b)  & Line       & $y=ax+b$    &  0.29   & 1.9    & 0.21    \\
\ref{fig:RDVsTTC}(b)  & Exponential& $y=ae^{bx}$ &  0.27   & 1.15   & 0.62    \\
\ref{fig:RDVsTTC}(b)  & Power      & $y=ax^{b}$  &  0.30   & 2.1    & 0.89    \\
\ref{fig:RDVsTTC}(c)  & Line       & $y=ax+b$    &  0.75   & -6.9   & 4.9     \\
\ref{fig:RDVsTTC}(c)  & Exponential& $y=ae^{bx}$ &  0.84   & 23     & -5.9    \\
\ref{fig:RDVsTTC}(c)  & Power      & $y=ax^{b}$  &  0.82   & 0.2    & -2.5    \\
\ref{fig:RDVsTTC}(d)  & Line       & $y=ax+b$    &  0.69   & -3.0   & 5.5     \\
\ref{fig:RDVsTTC}(d)  & Exponential& $y=ae^{bx}$ &  0.77   & 14.3   & -1.8    \\
\ref{fig:RDVsTTC}(d)  & Power      & $y=ax^{b}$  &  0.72   & 2.1    & -1.7    \\
\hline
\end{tabular}
\end{center}
\end{table}
\section{Conclusion and future works}
Integrating new users into current public spaces requires metrics for users' safety and comfort.
Time to collision is a prevalent metric in car traffic safety assessment.
We adapted the concept to define the perceived time to collision and experimentally evaluated it for e-scooter and pedestrian interactions.
The results show a strong correlation between the perceived time to collision and the reported discomfort when the agents see each other.
In addition, we calibrated functions of $T_p$ for online estimation of the discomfort.
Since the metric only uses relative velocity and position information, it is a viable candidate for neighboring people's discomfort estimation either in ADAS for e-scooters and PMVs or in robotics.
\par
For further research, we will focus on incorporating neighboring people's discomfort in the robot control algorithm.
The goal is to improve people's experiences on a futuristic sidewalk with PMVs and robots moving along with pedestrians.


\bibliography{ifacconf}             

\begin{thebibliography}{23}
\providecommand{\natexlab}[1]{#1}
\providecommand{\url}[1]{\texttt{#1}}
\providecommand{\urlprefix}{URL }
\expandafter\ifx\csname urlstyle\endcsname\relax
  \providecommand{\doi}[1]{doi:\discretionary{}{}{}#1}\else
  \providecommand{\doi}{doi:\discretionary{}{}{}\begingroup
  \urlstyle{rm}\Url}\fi

\bibitem[{Archer(2005)}]{Archer2005}
Archer, J. (2005).
\newblock \emph{{Indicators for traffic safety assessment and prediction and
  their application in micro-simulation modelling: A study of urban and
  suburban intersections}}.
\newblock PhD dissertation, KTH, Stockholm.

\bibitem[{Casella and Berger(2002)}]{RSquared}
Casella, G. and Berger, R.L. (2002).
\newblock \emph{Statistical inference (Second ed.)}.
\newblock Thomson Learning, Australia.

\bibitem[{Formosa et~al.(2022)Formosa, Quddus, Ison, and Timmis}]{FORMOSA}
Formosa, N., Quddus, M., Ison, S., and Timmis, A. (2022).
\newblock A new modeling approach for predicting vehicle-based safety threats.
\newblock \emph{IEEE Transactions on Intelligent Transportation Systems},
  23(10), 18175--18185.

\bibitem[{Gao and Huang(2022)}]{Rob_Nav_2022}
Gao, Y. and Huang, C.M. (2022).
\newblock Evaluation of socially-aware robot navigation.
\newblock \emph{Frontiers in Robotics and AI}, 420.

\bibitem[{Hasegawa et~al.(2018)Hasegawa, Dias, Iryo-Asano, and
  Nishiuchi}]{Hasegawa2018}
Hasegawa, Y., Dias, C., Iryo-Asano, M., and Nishiuchi, H. (2018).
\newblock Modeling pedestrians’ subjective danger perception toward personal
  mobility vehicles.
\newblock \emph{Transportation Research Part F: Traffic Psychology and
  Behaviour}, 56, 256--267.

\bibitem[{Hayward(1972)}]{Hayward1972}
Hayward, J.C. (1972).
\newblock Near miss determination through use of a scale of danger.
\newblock \emph{Pennsylvania State University University Park}.

\bibitem[{Hetherington et~al.(2021)Hetherington, Croft, and Van~der
  Loos}]{RobotExample}
Hetherington, N.J., Croft, E.A., and Van~der Loos, H.M. (2021).
\newblock Hey robot, which way are you going? nonverbal motion legibility cues
  for human-robot spatial interaction.
\newblock \emph{IEEE Robotics and Automation Letters}, 6(3), 5010--5015.

\bibitem[{Jang et~al.(2012)Jang, Choi, and Cho}]{Jang2012}
Jang, J.A., Choi, K., and Cho, H. (2012).
\newblock A fixed sensor-based intersection collision warning system in
  vulnerable line-of-sight and/or traffic-violation-prone environment.
\newblock \emph{IEEE Transactions on Intelligent Transportation Systems},
  13(4), 1880--1890.

\bibitem[{Jiang et~al.(2015)Jiang, Wang, and Bengler}]{Jiang2015}
Jiang, X., Wang, W., and Bengler, K. (2015).
\newblock Intercultural analyses of time-to-collision in vehicle–pedestrian
  conflict on an urban midblock crosswalk.
\newblock \emph{IEEE Transactions on Intelligent Transportation Systems},
  16(2), 1048--1053.

\bibitem[{Jo et~al.(2022)Jo, Jang, Ko, and Oh}]{Jo2022}
Jo, Y., Jang, J., Ko, J., and Oh, C. (2022).
\newblock An in-vehicle warning information provision strategy for v2v-based
  proactive traffic safety management.
\newblock \emph{IEEE Transactions on Intelligent Transportation Systems},
  23(10), 19387--19398.

\bibitem[{Jung et~al.(2020)Jung, Kim, Lee, Bang, Kim, and Kim}]{WhCExmple}
Jung, Y., Kim, Y., Lee, W.H., Bang, M.S., Kim, Y., and Kim, S. (2020).
\newblock Path planning algorithm for an autonomous electric wheelchair in
  hospitals.
\newblock \emph{IEEE Access}, 8, 208199--208213.

\bibitem[{Kazemzadeh et~al.(2020)Kazemzadeh, Laureshyn, Ronchi, D'Agostino, and
  Hiselius}]{E-BikeBehavior}
Kazemzadeh, K., Laureshyn, A., Ronchi, E., D'Agostino, C., and Hiselius, L.W.
  (2020).
\newblock Electric bike navigation behaviour in pedestrian crowds.
\newblock \emph{Travel Behaviour and Society}, 20, 114--121.

\bibitem[{Kilicarslan and Zheng(2019)}]{Kilicarslan2018}
Kilicarslan, M. and Zheng, J.Y. (2019).
\newblock Predict vehicle collision by ttc from motion using a single video
  camera.
\newblock \emph{IEEE Transactions on Intelligent Transportation Systems},
  20(2), 522--533.

\bibitem[{Liu et~al.(2022)Liu, Jafari, Shim, and Paley}]{T-ITS-I}
Liu, Y.C., Jafari, A., Shim, J.K., and Paley, D.A. (2022).
\newblock Dynamic modeling and simulation of electric scooter interactions with
  a pedestrian crowd using a social force model.
\newblock \emph{IEEE Transactions on Intelligent Transportation Systems},
  23(9), 16448--16461.

\bibitem[{Masaki and Motoi(2020)}]{Masaki2020}
Masaki, R. and Motoi, N. (2020).
\newblock Remote control method with force assist based on time to collision
  for mobile robot.
\newblock \emph{IEEE Open Journal of the Industrial Electronics Society}, 1,
  157--165.

\bibitem[{McGill et~al.(1978)McGill, Tukey, and Larsen}]{BoxChart}
McGill, R., Tukey, J.W., and Larsen, W.A. (1978).
\newblock Variations of box plots.
\newblock \emph{The American Statistician}, 32, 12--16.

\bibitem[{Salvini et~al.(2022)Salvini, Paez-Granados, and Billard}]{Saf_Con}
Salvini, P., Paez-Granados, D., and Billard, A. (2022).
\newblock Safety concerns emerging from robots navigating in crowded pedestrian
  areas.
\newblock \emph{International Journal of Social Robotics}, 14(2), 441--462.

\bibitem[{Satake et~al.(2013)Satake, Kanda, Glas, Imai, Ishiguro, and
  Hagita}]{App_Ped}
Satake, S., Kanda, T., Glas, D.F., Imai, M., Ishiguro, H., and Hagita, N.
  (2013).
\newblock A robot that approaches pedestrians.
\newblock \emph{IEEE Transactions on Robotics}, 29(2), 508--524.

\bibitem[{Shahriari and Biglarbegian(2022)}]{Shahriari2022}
Shahriari, M. and Biglarbegian, M. (2022).
\newblock Toward safer navigation of heterogeneous mobile robots in distributed
  scheme: A novel time-to-collision-based method.
\newblock \emph{IEEE Transactions on Cybernetics}, 52(9), 9302--9315.

\bibitem[{Shiomi et~al.(2014)Shiomi, Zanlungo, Hayashi, and Kanda}]{Shiomi2014}
Shiomi, M., Zanlungo, F., Hayashi, K., and Kanda, T. (2014).
\newblock Towards a socially acceptable collision avoidance for a mobile robot
  navigating among pedestrians using a pedestrian model.
\newblock \emph{International Journal of Social Robotics}, 6(3), 443--455.

\bibitem[{Tak et~al.(2015)Tak, Kim, and Yeo}]{Tak2015}
Tak, S., Kim, S., and Yeo, H. (2015).
\newblock Development of a deceleration-based surrogate safety measure for
  rear-end collision risk.
\newblock \emph{IEEE Transactions on Intelligent Transportation Systems},
  16(5), 2435--2445.

\bibitem[{Zhang et~al.(2022)Zhang, Yuan, Chu, Huang, Ding, Yuan, and
  Chen}]{Zhang2022}
Zhang, L., Yuan, K., Chu, H., Huang, Y., Ding, H., Yuan, J., and Chen, H.
  (2022).
\newblock Pedestrian collision risk assessment based on state estimation and
  motion prediction.
\newblock \emph{IEEE Transactions on Vehicular Technology}, 71(1), 98--111.

\bibitem[{Zhang et~al.(2012)Zhang, Yao, Qiu, Peng, and Zhang}]{Zhang2012}
Zhang, Y., Yao, D., Qiu, T.Z., Peng, L., and Zhang, Y. (2012).
\newblock Pedestrian safety analysis in mixed traffic conditions using video
  data.
\newblock \emph{IEEE Transactions on Intelligent Transportation Systems},
  13(4), 1832--1844.

\end{thebibliography}


\begin{thebibliography}{4}
\providecommand{\natexlab}[1]{#1}
\providecommand{\url}[1]{\texttt{#1}}
\providecommand{\urlprefix}{URL }
\expandafter\ifx\csname urlstyle\endcsname\relax
  \providecommand{\doi}[1]{doi:\discretionary{}{}{}#1}\else
  \providecommand{\doi}{doi:\discretionary{}{}{}\begingroup
  \urlstyle{rm}\Url}\fi

\bibitem[{Able(1956)}]{Abl:56}
Able, B. (1956).
\newblock Nucleic acid content of microscope.
\newblock \emph{Nature}, 135, 7--9.

\bibitem[{Able et~al.(1954)Able, Tagg, and Rush}]{AbTaRu:54}
Able, B., Tagg, R., and Rush, M. (1954).
\newblock Enzyme-catalyzed cellular transanimations.
\newblock In A.~Round (ed.), \emph{Advances in Enzymology}, volume~2, 125--247.
  Academic Press, New York, 3rd edition.

\bibitem[{Keohane(1958)}]{Keo:58}
Keohane, R. (1958).
\newblock \emph{Power and Interdependence: World Politics in Transitions}.
\newblock Little, Brown \& Co., Boston.

\bibitem[{Powers(1985)}]{Pow:85}
Powers, T. (1985).
\newblock Is there a way out?
\newblock \emph{Harpers}, 35--47.

\end{thebibliography}
\end{document}